\documentclass[aps,pra,epsfigure,notitlepage,twocolumn,longbibliography,superscriptaddress]{revtex4-1}

\usepackage{bm} 
\usepackage{graphicx}
\usepackage{amsmath}    
\usepackage{latexsym}
\usepackage{amsfonts}   
\usepackage{amssymb}
\usepackage{comment}
\usepackage{array}      
\usepackage{epsfig}
\usepackage{txfonts}
\usepackage{xcolor}
\usepackage[colorlinks=true,linkcolor=blue,urlcolor=blue,citecolor=blue,pdfusetitle]{hyperref}
\usepackage{hyperref}
\usepackage{ulem}

\newcommand{\braket}[2]{\langle#1|#2\rangle}

\newcommand{\tr}{{\mbox{tr}}}

\usepackage{ulem}

\usepackage{tikz}
\usetikzlibrary{calc}

\begin{document}
\title{Non-Gaussian work statistics at finite-time driving}
\author{Krissia Zawadzki$^*$}
\affiliation{School of Physics, Trinity College Dublin, College Green, Dublin 2, Ireland}
\altaffiliation[]{These authors contributed equally to this work.}
\author{Anthony Kiely$^*$}
\affiliation{School of Physics, University College Dublin, Belfield, Dublin 4, Ireland}
\affiliation{Centre for Quantum Engineering, Science, and Technology,
University College Dublin, Belfield, Dublin 4, Ireland}
\author{Gabriel T. Landi}
\affiliation{Instituto de F\'isica da Universidade de S\~ao Paulo,  05314-970 S\~ao Paulo, Brazil.}
\affiliation{School of Physics, Trinity College Dublin, College Green, Dublin 2, Ireland}
\author{Steve Campbell}
\affiliation{School of Physics, University College Dublin, Belfield, Dublin 4, Ireland}
\affiliation{Centre for Quantum Engineering, Science, and Technology,
University College Dublin, Belfield, Dublin 4, Ireland}

\begin{abstract}
We study properties of the work distribution of a many-body system driven through a quantum phase transition in finite time. 
We focus on the non-Gaussianity of the distribution, which we characterize through two quantitative metrics: skewness and negentropy.
In particular, we focus on the quantum Ising model and show that a finite duration of the ramp enhances the non-Gaussianity of the distribution for a finite size system. By examining the characteristics of the full distribution, we observe that there is a clear intermediate regime between the sudden quench and adiabatic limits, where the distribution becomes increasingly skewed. 
\end{abstract}
\maketitle

\section{Introduction}

The work that is injected into, or extracted from, a microscopic system may fluctuate significantly.
These fluctuations are not only relevant from a practical perspective, but they also encode fundamental results concerning the second law of thermodynamics~\cite{Jarzynski1997,Crooks1998}. Over the last two decades this has motivated a flurry of interest in understanding and characterizing work fluctuations. In the simplest scenario, the problem can be framed as that of a system driven externally by a time-dependent work protocol, which causes its Hamiltonian to be time dependent for a certain duration $\tau$. 
The work distribution $P(W)$ is then obtained by measuring the system's energy before and after the drive~\cite{Talkner2007}. 
In particular, a problem that has received considerable attention is the work statistics of quantum critical systems. It dates back to the early days of quantum thermodynamics \cite{prl_silva}, and is motivated by the goal of understanding how far from equilibrium the system goes when driven across its critical point \cite{Talkner2007, WorkDistPRL, john_chapter_book}. 

In this regard, the duration $\tau$ of the drive plays a crucial role. 
Extremely rapid dynamics (the so-called ``sudden quench regime'') tend to produce a large number of excitations and have been the focus of several recent works~\cite{prl_silva, CampbellPRB, prx_fusco, ArgentinianPRE, 2017_quan, CampbellPRL2020, LandiPRR, CampbellPRE}. Interestingly, it has been observed that in this regime $P(W)$ for a variety of closed many-body systems tends to a Gaussian \cite{Sotiriadis-PRE, Marino_PRB_TFIM_Gaussian, Chenu2018SciRep,AndreiPRB2019, ArgentinianPRE,Abeling2016}, and therefore most papers have focused only on the first and second moments.
At the other extreme, for slow protocols the dynamics will be effectively adiabatic. 
This regime has also received a lot of attention because it is at the core of thermal cycles and serves as a basis to achieve protocols in which the system remains close to equilibrium \cite{Scandi2020, Miller2021}. Furthermore, in contrast to sudden quenches, in this case it has been shown that quantum coherences lead to non-Gaussian work distributions~\cite{Miller2019,Scandi2020}.

In between the sudden quench and slow driving regimes,  finite time protocols can give rise to highly non trivial behaviors, such as those captured by the Kibble-Zurek mechanism~\cite{Kibble1976, Zurek1985,Damski2005,Zurek2005,PueblaPRR,EspositoPRL2020}. Indeed, as discussed in Ref.~\cite{KrissaPRR}, the interplay between the time-dependent driving and the characteristic energy scales implies that contributions of different excitations change dynamically. In practice, the speed of the external driving determines which states are accessible at the beginning and the end of the process. This is encoded in the conditional probabilities, $p_{m|n}$, of the system being in state $m$ given that it was initially in state $n$. 
Modifying the structure of accessible states, and consequently the possible transitions $p_{m|n}$, has a strong impact on the statistics of the work distribution that extends beyond the first two moments.
Recently, it was further shown that the statistics of the work distribution exhibit a universal scaling analogous to the Kibble-Zurek mechanism (KZM) scaling of topological defects~\cite{Dong2019,EspositoPRL2020}. 
Moreover, finite time effects have been used to refine bounds on probabilistic violations of the second law in the work distribution~\cite{Miller2022}.

In this work, we add to this endeavor by characterzing the work distribution in terms of its Gaussianity. Focusing on the transverse field Ising chain, we carry out a detailed analysis of the skewness of the distribution as a function of the chain size and the drive time. 
We show that there is a highly non-trivial interplay between both, which precisely captures the transition from fast to slow driving. 
In addition, we study the relative entropy of non-Gaussianity, also known as negentropy. We explain some subtleties in employing this metric for discrete distributions, but notwithstanding demonstrate that it does provide a useful quantifier of non-Gaussianity, which goes a step beyond the skewness by also incorporating information related to higher order moments of the distribution. In particular, both methods establish that there is a clear intermediate regime between the sudden quench and adiabatic limits, where the distribution tends to become increasingly non-Gaussian.

\section{Work statistics in driven quantum systems}
\subsection{Distribution, moments and cumulants}
We consider an isolated quantum system, prepared in an initial state $\rho$. At time $t=0$ the system is driven by a time-dependent  Hamiltonian $H_t= \sum_n E_n^t |n_t\rangle\langle n_t\rangle$ for a total time $\tau$. 
The final state of the system is thus $\rho_\tau =  U \rho U^\dagger$, where
$U=\mathcal{T}\exp\left(-\frac{i}{\hbar}\int_0^\tau ds H_s\right)$  and $\mathcal{T}$ is the Dyson time-ordering operator.
We are interested in the work distribution as the system Hamiltonian is changed from $H_0$ to $H_\tau$. This can be obtained following the standard two-point measurement~\cite{Talkner2007}, which consists of projectively measuring the energy of the system before and after the drive. 
For isolated systems, as there is no surrounding environment, defining the work extracted in terms of changes in the system's energy is reasonable since, by definition, there is no heat exchange. Even in the case of mixed states, if the dynamics is unitary then energetic variations due to changes in the Hamiltonian can be solely identified as work~\footnote{We refer the reader to alternative definitions of work beyond the two point measurement scheme in for example Refs.~\cite{RAlicki_1979, Frenzel_PhysRevE.90.052136, Dahlsten_2017, PhysRevE.94.010103, Alipour_PhysRevA.105.L040201}.} The corresponding work will then be one of the possible energy differences between the final and initial Hamiltonian, $W = E_m^\tau - E_n^0$, which occur with probability 
\begin{equation}\label{PW}
    P(W) = \sum\limits_{n,m}^N \langle n_0 |\rho| n_0\rangle~p_{m|n}~\delta\left[W - (E_m^\tau - E_n^0)\right]
\end{equation}
where $p_{m|n}=|\langle m_\tau | U| n_0 \rangle|^2$. Note that in the adiabatic limit $p_{m|n}=\delta_{m,n}$ and for sudden quenches $p_{m|n}=|\braket{m_\tau}{n_0}|^2$.

The expression simplifies when the initial state is an eigenstate of $H_0$. For instance, if $\rho$ is  the ground state, $\rho = |{\rm gs}_0\rangle\langle {\rm gs}_0|$, the work distribution becomes 
\begin{equation}\label{PW_gs}
    P(W) = \sum\limits_{m} |\langle m_\tau | U| {\rm gs}_0 \rangle|^2~\delta\left[W - \left(E_m^\tau - E_{\rm gs}^0\right)\right].
\end{equation}
The work distribution in this case is a local density of states, which is essentially probing the spectrum $E_m^\tau$ of the final Hamiltonian, with weights given by $|\langle m_\tau | U| {\rm gs}_0 \rangle|^2$.

The characteristic function (CF) of the work distribution Eq.~\eqref{PW} is given by
\begin{equation}\label{CF}
    G(u) := \langle e^{i u W} \rangle = \tr\Big(U^\dagger e^{i u H_\tau} U e^{-i u H_0} \rho_d\Big),
\end{equation}
where $\rho_d = \sum_n |n_0\rangle\langle n_0|\rho|n_0\rangle\langle n_0|$ is the initial state, dephased in the basis of $H_0$. 
From $G(u)$ the moments can be computed as $\langle W^n \rangle = i^{-n} G^{(n)}(0)$. 
Carrying out the expansion we find 
\begin{equation}\label{moments}
    \langle W^n \rangle = \sum\limits_{k=0}^n \binom{n}{k} (-1)^{n-k} \tr\Big\{ (U^\dagger H_\tau U)^k H_0^{n-k} \rho_d\Big\}.
\end{equation}
This provides a convenient expression to compute all moments, without having to actually construct $P(W)$.
From the CF one also builds the cumulant generating function (CGF)  $C(u)=\ln G(u)$. A series expansion then yields the cumulants as $\kappa_m=i^{-m} C^{(m)}(0)$. 
The first cumulant is the mean, $\kappa_1 \equiv \mu$, and the second is the variance $\kappa_2 \equiv \sigma^2$. 
The first three cumulants coincide with the corresponding central moments, $\mu_n := \langle (W-\langle W \rangle)^n \rangle$, but this is no longer true for $n>3$. For instance, 
$\kappa_4 = \mu_4 -3 \mu_2^2$, and so on. 

The cumulants are particularly convenient to characterize the dependence of the work distribution with the size $L$ of the system (e.g., the number of lattice sites, or the number of particles).
Depending on the structure of the Hamiltonian, and the type of drive employed, the dependence with $L$ may vary significantly. 
A useful reference, for comparison, is when the work is associated to $L$ independent sources. This happens, for instance, if the system is composed of $L$ non interacting particles. This is also naturally expected in models which are exactly solvable via a mapping to free fermions~\footnote{We remark that for truly strongly correlated systems this will necessarily break down~\cite{ToAppearKrissia}.}.

In this case the statistics become extensive and the central limit theorem applies. 
As a consequence, all cumulants of $W$  grow linearly with $L$:
\begin{equation}\label{standard_scenario}
\kappa_n(W) \sim L.    
\end{equation}
If one defines a rescaled work variable $w = W/\sqrt{L}$, the corresponding cumulants will then scale as
\begin{equation}
    \kappa_n(w)=(1/\sqrt{L})^n\kappa_n(W) \sim L^{1-\frac{n}{2}}.
\end{equation}
That is
\begin{align*}
    \kappa_1(w) &\sim \sqrt{L}, 
    &\kappa_2(w) &\sim 1, 
    \\[0.2cm]
    \kappa_3(w) &\sim 1/\sqrt{L}, 
    &\kappa_4(w) &\sim 1/L.
\end{align*}
For large $L$ all cumulants with $n\geqslant 3$ tend to be suppressed, and the work distribution will therefore tend to a Gaussian.

\subsection{Measures of non-Gaussianity and negentropy}
\label{measures}
The above discussion highlights the Gaussianity (or lack thereof) of $P(W)$ as an interesting feature that may provide insight in characterizing  different models and/or dynamical regimes~\cite{Genoni2010,Takagi2018,Albarelli2018,Zhuang2018,Santalla2017}. In this section we discuss measures to quantify the degree of non-Gaussianity. 
The simplest approach is to analyze the cumulants $\kappa_3(W)$, $\kappa_4(W)$, etc. 
It is more convenient to work with the dimensionless quantities 
\begin{equation}
    \lambda_m := \frac{\kappa_m(W)}{\kappa_2(W)^{m/2}}.
\end{equation}
The skewness is 
$\lambda_3 = \kappa_3/\kappa_2^{3/2} = \kappa_3/\sigma^3$, while $\lambda_4 = \kappa_4/\kappa_2^2 = \kappa_4/\sigma^4$ is related to the kurtosis as $\lambda_4 + 3$. 
In the standard scenario of Eq.~\eqref{standard_scenario}, they scale according to 
\begin{equation}\label{standard_scenario_lambda}
\lambda_m \sim L^{1-m/2},    
\end{equation}
such that a vanishing $\lambda_m$ ($m\geqslant 3$) can be used to quantify wheter the work distribution is tending to a Gaussian. Many normality tests used in statistics, such as the Jarque-Bera~\cite{Jarque1980} or D'Agostino's $K$-squared tests~\cite{D_AGOSTINO1970}, are in fact based on $\lambda_3$ and~$\lambda_4$. 

Alternatively, one may employ an information-theoretic approach, reminiscent of quantum resource theories. A monotone of non-Gaussianity can be built using the relative entropy (Kullback-Leibler divergence) between $P(W)$ and a corresponding Gaussian distribution
$P_G(W)=\frac{1}{\sqrt{2 \pi}\sigma} e^{-(W-\mu)^2/2\sigma^2}$, with the same mean $\mu$ and variance $\sigma^2$ as $P(W)$. 
For a given $P(W)$, the quantifier is therefore defined as
\begin{eqnarray}
\label{eq:distance_Gaussian}
J(P) := D\left(P || P_G\right) &=& \int_{-\infty}^{\infty} P(W) \ln \left[\frac{P(W)}{P_G(W)}\right] dW.
\end{eqnarray}
This quantity is usually termed negentropy.
The integral can actually be carried out further, leading to
\begin{equation}\label{negentropy2}
    J(P) = S(P_G)-S(P) = \ln\left(\sqrt{2\pi e} \sigma\right) - S(P),
\end{equation}
where 
\begin{equation}\label{differential_entropy}
S(p)=-\int_{-\infty}^{\infty} p(x) \ln p(x) \, dx,    
\end{equation}
is the differential entropy of $p(x)$. 
The concept of negentropy can also be extended to quantum states, in the context of resource theories of non-Gaussianity~\cite{Genoni2010,Takagi2018,Albarelli2018,Zhuang2018}. In that case, the meaningful entropic quantities are instead the von Neumann entropy and the quantum relative entropy. 
Despite the underlying quantum system, our interest here is in the classical version of the negentropy, since our object of study is the classical probability distribution $P(W)$.

For continuous probability distributions, the negentropy is a very good quantifier of non-Gaussianity: it is non-negative and vanishes if and only if the distribution is Gaussian. 
In our case, however, an issue arises concerning the discrete nature of Eq.~\eqref{PW}. 
Namely, because the support [the points $W$ where $P(W) \neq 0$] is not compact, the corresponding differential entropy~\eqref{differential_entropy} is not well defined. In fact, this is a famous issue found by Shannon, and later resolved by Jaynes using the concept of a limiting density of discrete points~\cite{Jaynes1968}. However, unfortunately there is no unambiguous way of addressing it. 

One approach, which directly connects with the Jarque-Bera test~\cite{Jarque1980}, is to approximate $P(W)$ by a continuous smooth distribution using the Gram-Chalier series~\cite{Wallace1958}, 
which effectively provides a systematic set of corrections to a Gaussian distribution in terms of $\lambda_3,\lambda_4,\ldots$. 
Stopping at fourth order, the approximate distribution will have the form 
\begin{align}
    P_{\rm app}(W)&= P_G(W)\left[ 1+f(v)\right], \label{GCE_Series} \\[0.2cm]
    f(v) &= \frac{\lambda_3}{3!}\mathit{H}_3\left(v\right)+\frac{\lambda_4}{4!}\mathit{H}_4\left(v\right),
\end{align}
where $\mathit{H}_n(x)$ are the probabilist's Hermite polynomials.
For $\lambda_3 \sim L^{-1/2}$ and $\lambda_4 \sim L^{-1}$, the correction terms become vanishingly small with increasing $L$.
This expansion applies not only to the standard scenario of Eqs.~\eqref{standard_scenario} and~\eqref{standard_scenario_lambda}, but also to any scenario where cumulants of order 3 or higher are small. This includes, for instance, sub/super extensive dependences in $L$;
that is, $\kappa_m \sim L^q$ for $q \lessgtr1$, which then implies $\lambda_m \sim L^{q(1-m/2)}$. 
We remark that there are circumstances in which this approach may not provide a good approximation for the distribution, particularly when the distributions are not smooth as is the case for certain chaotic systems~\cite{Grabarits2022classical, Mata-PhysRevE.95.050102, ArgentinianPRE} and for systems with a fractal spectrum, such as the Aubry-Andre-Harper model~\cite{Anthony_2022}.

Following Ref.~\cite{Hulle2005}, when $f$ is small we can series expand 
\begin{eqnarray}
P_{\rm app} \ln P_{\rm app} 
&\approx &  P_G\left[ (1+f) \ln P_G + f+\frac{f^2}{2}\right].
\end{eqnarray}
The negentropy~\eqref{negentropy2} can then be written as
\begin{equation*}
J(P_{\rm app}) 
=  \left[1-\ln\left(\sqrt{2 \pi}\sigma\right)\right]\left\langle f\right\rangle_G -\frac{1}{2} \left\langle v^2 f\right\rangle_G +\frac{1}{2}\left\langle f^2\right\rangle_G,
\end{equation*}
where $\langle \ldots \rangle_G$ denotes expectation values over $P_G(W)$. 
Finally, using the orthogonality relations of the Hermite polynomials, we find that
$\left\langle f\right\rangle_G=0$ and $\left\langle v^2 f\right\rangle_G=0$.
This then leads us to
$J(P_{\rm app}) = \frac{1}{2} \langle f^2\rangle_G \nonumber $ or, more explicitly, 
\begin{equation}
J(P) = \frac{1}{2}\left[ \frac{\lambda_3^2}{3!} +\frac{\lambda_4^2}{4!} \right].
\label{negentropy_approx}
\end{equation}
This result, which also coincides with the Jarque-Bera test~\cite{Jarque1980} used in statistics, provides us with a clear measure of deviations from Gaussianity in terms of  $\lambda_3$ and $\lambda_4$. 
In principle, one could also extend this procedure and consider higher order cumulants. 
However, in practice this is typically not very useful since higher order cumulants are extremely sensitive to numerical errors.

\begin{figure*}[t!]
\begin{center}
	\includegraphics[width=2\columnwidth]{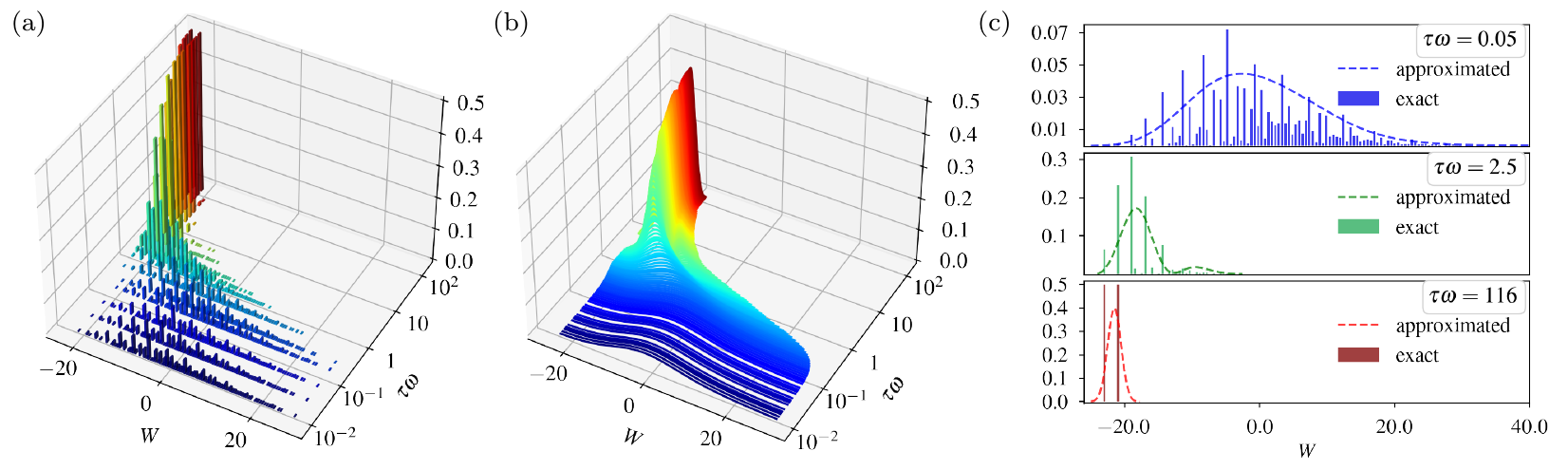}
        


\end{center}
\caption{(a) Full work distribution, Eq.~\eqref{PW_gs}, for an Ising spin chain with $L\!=\!20$ as a function of quench duration. (b) Approximation to the full distribution constructed using the first four cumulants. (c) Comparison between the exact discrete (bars) and approximate continuous (dashed lines) distributions for fixed values of ramp duration, $\tau$. In all panels we consider a linear ramp, Eq.~\eqref{linramp}, with the system starting in the ground state and $g_0\!=\!0$.}
 \label{DistED_Approx}
\end{figure*}
\section{Numerical Results}
We now apply the above ideas to a paradigmatic model for critical quantum systems, namely the transverse field Ising model (TFIM) consisting of $L$ sites described by Pauli matrices $\sigma_i^{x,y,z}$, with Hamiltonian 
\begin{equation}
    H_t=- \hbar \omega\sum_{i=1}^{L}\left[g(t) \,\sigma_i^x+\sigma_i^z\sigma_{i+1}^z\right].
\end{equation}
We assume $L$ is even, and use periodic boundary conditions $\sigma_{L+1}^{x,y,z}\!=\!\sigma_1^{x,y,z}$.
Moreover, we henceforth set $\hbar\omega=1$, thus fixing the scales of energy. 

The model is well known to exhibit a second-order quantum phase transition at $g_c\!\!=\!\!1$ \cite{prl_silva, Dorner-PRL-work-TFIM, Heyl-PRL-work-TFIM, EspositoPRL2020, Abeling2016, Russomanno2015, ParaanPRE, XuPRA2018, prx_fusco, LandiPRR, Scandi2020}. It has been demonstrated theoretically that, when the system is driven at finite-time through its critical point, the distribution of topological defects follows an universal scaling compatible with the Kibble-Zurek mechanism \cite{KZM_Adolfo2018}. The power laws followed by the first three cumulants were tested experimentally with trapped ions \cite{KZM-PhysRevResearch.2.033369} and quantum annealers \cite{KZM-Cui2020experimentally,KZM-Dwave-king2022coherent}.

We explore the properties of the full work distribution for a system driven symmetrically through the critical point by a linearly varying drive
\begin{equation}
\label{linramp}
    g(t)\!=\!g_0+2(g_c-g_0) t/\tau
\end{equation} 
where $\tau$ is the duration but remark that we expect our results to be qualitatively similar for other suitable choices of ramp. We assume that the system begins in the ground state manifold for $g_0\!=\!0$ and we calculate the full work distribution, Eq.~\eqref{PW_gs}, of the TFIM via exact diagonalization (ED). Translational and spin inversion symmetries allow us to solve systems with up to 20 spins since the ground state only connects to states sharing the same conserved quantum numbers~\cite{Sacco2022}.

\begin{figure*}[t]
\begin{center}
\includegraphics[width=0.95\linewidth]{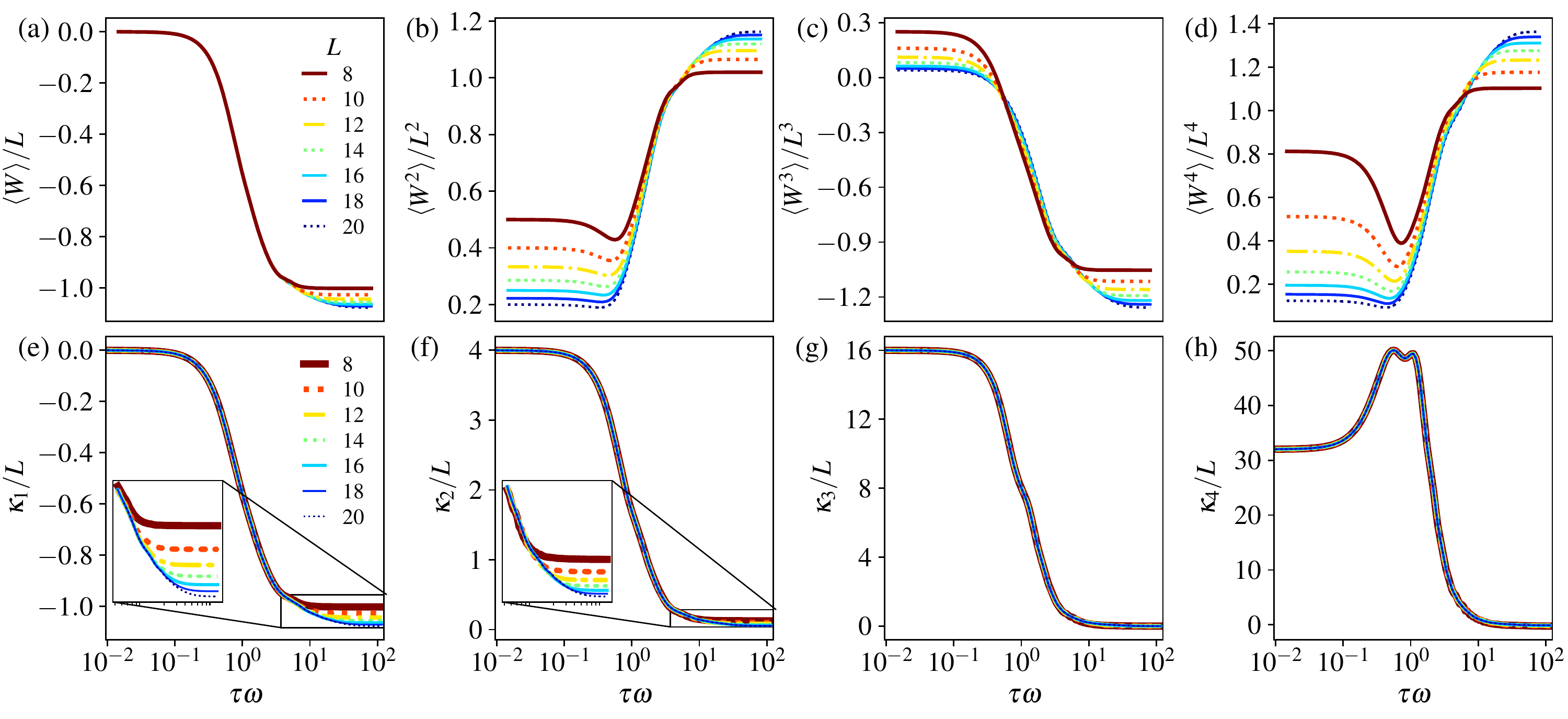}
\end{center}
\caption{(a--d) Moments of work distribution for Ising chains of size $L=8$ (dark red solid line), $L=10$ (orange dotted line),  $L=12$ (yellow dot dashed line), $L=14$ (lime green dotted line), $L=16$ (cyan solid line), $L=18$ (blue solid line), and $L=20$ (indigo dotted line). (e--h) Cumulants of work distribution for Ising chains of size $L=8$ (dark red solid line), $L=10$ (orange dotted line),  $L=12$ (yellow dot dashed line), $L=14$ (lime green dotted line), $L=16$ (cyan solid line), $L=18$ (blue solid line), and $L=20$ (indigo dotted line). In all panels we consider a linear ramp, Eq.~\eqref{linramp}, with the system starting in the ground state and $g_0\!=\!0$.}
 \label{momentsED}
\end{figure*}


\begin{figure}[t]
\begin{center}
\includegraphics[width=0.99\columnwidth]{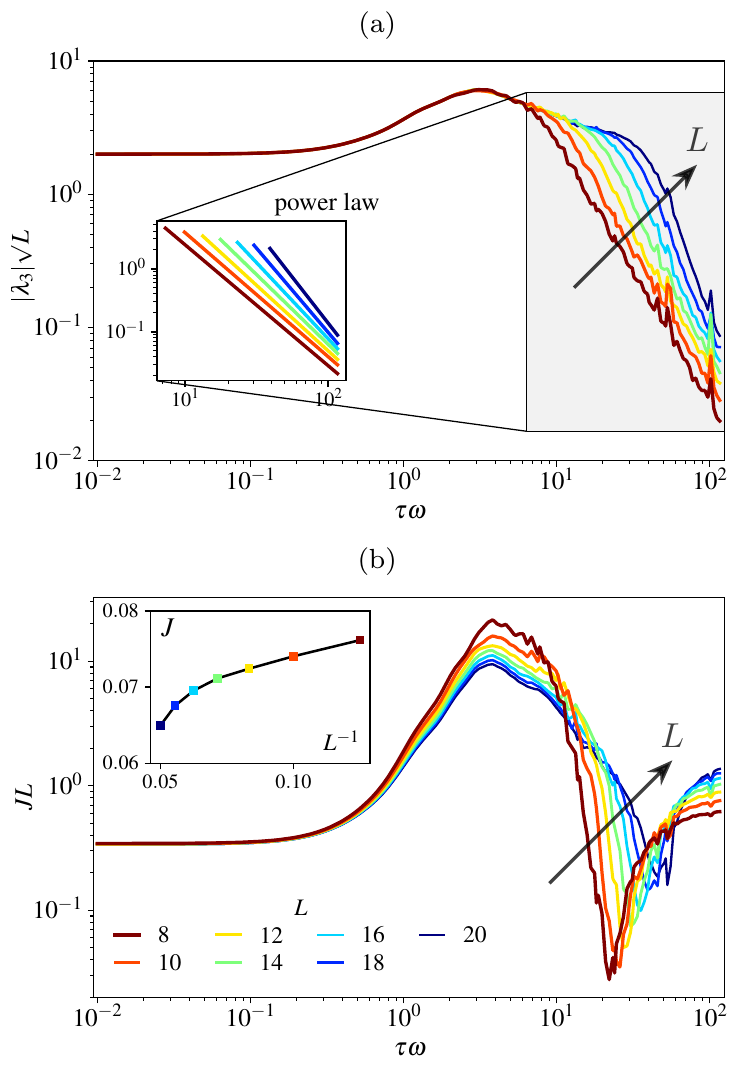}

\end{center}
\caption{Measures of non-Gaussianity of the work distribution for the linear ramp protocol. (a) Skewness and (b) negentropy for finite Ising chains. The inset in (a) shows a linear fit of each curve in the region where a power law is observed. The inset in (b) displays the unscaled, long-time values which the negentropy reaches near $\tau \sim 10^2$. In the quasi adiabatic regime, systems with different size become distinguishable: for $\tau \omega > 50$, the curve associated with the smallest system $L=8$ is at the bottom of the curves set, while the largest $L=20$ is at the top.}
 \label{SkewNegentropyED}
\end{figure}

The initial condition $g_0\!=\!0$ makes the ground state twofold degenerate, with each state related by spin inversion symmetry in the $z$ direction. In an adiabatic evolution, this degeneracy implies a splitting of the transition probabilities into two peaks with equal weight $p_{m|n}\! =\! 1/2$. In the limit of a sudden quench, when many excitations are created, these two ground states can overlap with a large number of eigenstates of the final Hamiltonian. While quantitative differences will occur for other choices of initial states, particularly if one restricts to a specific symmetry sector, the qualitative behaviors discussed remain largely unaffected.

In Fig.~\ref{DistED_Approx}(a) we show the exact work distribution for a $L\!=\!20$ site chain for the two-fold degenerate initial state. For $\tau \omega\! \ll \! 1$, $P(W)$ resembles a Gaussian: it is widely spread throughout the full spectrum, with the largest contributions centered at the average $\langle W \rangle$. For such fast, but manifestly not instantaneous, ramps we see that the distribution is nevertheless invariant. This indicates that the sudden quench approximation remains valid even for fast finite ramps. Conversely, for very slow protocols ($\tau \omega \! \rightarrow \! \infty$), the work distribution splits into two peaks, each one associated with the two ground states. An intermediate regime emerges for $1 \!\lesssim\! \tau \omega \! \lesssim \! 10$, for which the work distribution appears more involved and does not peak at the adiabatic ground states. In this regime, the skewness of $P(W)$ is enhanced towards negative values of the work.

From the full distribution we determine the first four moments $\langle W^m \rangle$ and cumulants $\kappa_m$ ($m=1,2,3,4$) as a function of the duration of the ramp $\tau$, for finite systems with sizes from $L\!=\!8$ to $L\!=\!20$. In Fig.~\ref{momentsED}, the first row shows the scaled moments $\langle (W/L)^m \rangle$, while the second  shows the scaled cumulants $\kappa_m / L$. In both, we observe a crossover between $10^{-1} \! <\! \tau \omega\! < \! 10 $, which separates the sudden quench and the (quasi)adiabatic regimes. Note that the scaling $(W/L)^k$ is non-linear, and the moments are sensitive to the system size. Regions where the cumulants deviate from extensive scaling (shown in the inset of Fig.~\ref{momentsED} (e) and (f)) are due to finite size effects. While the first three scaled cumulants display a similar behavior, the fourth shows a peak centered at $\tau \omega\! =\! 1$. The fluctuations, given by the second cumulant, are amplified for a fast driving due to the spread of the spectral weight of the instantaneous wave-function through the final states.

From the first four cumulants of the exact data in Fig.~\ref{DistED_Approx}(a), we can build an approximated distribution for $P(W)$ using the Gram–Charlier series in Eq.~\eqref{GCE_Series} with the first four cumulants. The result is shown in Fig.~\ref{DistED_Approx}(b). The structure of the exact distribution is preserved: close to a sudden quench, one can recover a nearly Gaussian profile; while at intermediate driving speeds, a skewed distribution emerges before then converging to a single sharp peak in the adiabatic limit. A comparison between the exact and approximated distributions for $\tau \omega \! \approx\! 0.05,~2.5,~116 $ is shown panel Fig.~\ref{DistED_Approx}(c). The accuracy of this approximation is expected to improve with increasing system size.

Turning to the characterisitics of the distributions, we quantitatively examine deviations of the work distribution from a Gaussian and its dependence on the speed of the external drive employing the metrics outlined in Sec.~\ref{measures}. To this aim, in Fig.~\ref{SkewNegentropyED}(a) we inspect the skewness, $\lambda_3$, of the full distribution calculated via ED as a function of $\tau \omega$ for different $L$. As remarked previously, given that all cumulants of a normal distribution vanish after second order, the skewness provides a readily accessible figure of merit to understand the effects of the finite-time dynamics in the work distribution. 
From Fig.~\ref{SkewNegentropyED}(a) we observe that in the fast driving regime the skewness is constant and scales as $1/\sqrt{L}$. At short times, we see that there is a residual value of $\lambda_3$, which in effect quantifies the smallest distance between the actual distribution and a perfect Gaussian for a finite system after a sudden quench. Notice that since the skewness scales inversely with the system size, it is clear that in the thermodynamic limit, one would expect $\lambda_3 \rightarrow 0$ for the Ising model. As we increase the ramp duration, the skewness increases for intermediate driving  times before peaking around $\tau \omega \!\sim\! 3$, after which it decreases according to a power law in $\tau \omega$. We see a clear universal behavior for the sudden quench and intermediate regimes, while features due to finite size effects become evident as the ramp approaches the adiabatic limit; the onset of the power law decay emerges at progressively longer ramp durations for larger system sizes due to the fact that the ground state energy gap decreases as $L^{-1}$. Thus, the time  scales to reach adiabaticity necessarily increase, approaching infinity as $L\!\to\!\infty$.

In Fig.~\ref{SkewNegentropyED}(b) we plot the scaled negentropy, $J L$, as a function of $\omega \tau$ for different system sizes. We immediately see some qualitative similarities with the skewness. In particular, the negentropy of the approximate distribution is constant for fast dynamics, where the sudden quench approximation is valid. However, in contrast with the skewness, the negentropy asymptotically approaches a fixed value for a given system size in the adiabatic limit, c.f. the inset of Fig.~\ref{SkewNegentropyED}(b) shows the (unscaled) values attained at $\tau\sim 10^2$. Once again, this can be understood from the distribution tending towards two equally likely values of work for the given choice of initial state, and in this limit the negentropy tends to $J \rightarrow 1/12$. Note that there is a slower convergence towards this value for larger system sizes as the decreased energy gap results in a longer timescale required for adiabaticity, as can be inferred from the inset of Fig.~\ref{SkewNegentropyED}(b) where $N\!=\!8$ is already close to the asymptotic value while larger systems clearly require longer timescales. In the intermediate regime we see that the negentropy captures the increasingly non-Gaussian characteristics of the full distribution, similarly peaking for a finite drive but now also exhibiting a sharp minimum which appears to signal the transition toward the adiabatic regime and arises due to contributions coming from the fourth cumulant.

\section{Conclusions}
We have examined how characteristics of the work distribution are dependent on the rate at which a system is driven through a quantum phase transition. We have established that there is an interesting trade-off between the (lack of) Gaussianity of a distribution and speed with which it is ramped. Higher order moments of the distribution were shown to be useful indicators of different dynamical regimes. While the first and second moments (corresponding to the mean and variance) are typically the most studied, in line with Ref.~\cite{KrissaPRR} we have shown the the skewness provides valuable information about the response of the system to the ramp protocol. 

In particular, we have developed a general framework to assess the Gaussianity of the work distribution in terms of the negentropy. This can be approximated by a series expansion of cumulants providing a simple means of calculation, however, we stress that due to the discreteness of the full distribution for finite sizes, systematic errors can emerge since the approximation attempts to resolve the fine structure of the distribution. These errors can be more critical in scenarios in which the system's spectrum and its level statistics constrain the set of accessible energy transitions to a few. For this reason, we have demonstrated that restricting to the first four cumulants is sufficient to obtain a very good approximation, while largely avoiding such pathological issues.

While our results are demonstrated for the Ising model, we expect that qualitatively similar behaviors would be exhibited in other models that can be mapped to free-fermions. Future work could go beyond the scenario presented here, e.g. considering thermal initial states, the effect of non-integrability, non-linear ramps, or more carefully assessing the role that non-Gaussian features play in the dynamics of quantum systems~\cite{Miller2019, HarryPRL2020}. Furthermore, the characterisations of non-Gaussianity that we have proposed requires knowledge of only a few cumulants and hence could be applicable to settings where cumulants can be easily calculated, but the full distribution is difficult to access.
\vspace{-0.1cm}
\acknowledgements
 A.K. and S.C acknowledge support from the Science Foundation Ireland Starting Investigator Research Grant ``SpeedDemon" No. 18/SIRG/5508. K.Z. acknowledges the European Research Council Starting Grant ODYSSEY (G. A. 758403) for financial support and Javier Laguna for insightful comments. S.C. acknowledges the John Templeton Foundation (Grant ID 62422). The package QuSpin~\cite{QuSpin} was employed in the ED calculations.

\bibliography{Finite_Time_WD_refs}

\end{document}